\documentclass[%
reprint,
superscriptaddress,
nofootinbib,
amsmath,amssymb,
aps,
prx,
]{revtex4-1}

\usepackage{amstext}
\usepackage{graphicx}
\usepackage{subcaption}
\usepackage{color}
\usepackage{soul}
\usepackage[font=small, labelfont=bf]{caption}
\usepackage{MnSymbol}
\usepackage{subfiles}
\usepackage{bm}
\usepackage{hyperref}
\usepackage{layouts}
\usepackage{enumitem}

\usepackage{graphicx} 

\begin{document}

\title{From Statistical Physics to Social Sciences:\\ The Pitfalls of Multi-disciplinarity}
\author{Jean-Philippe Bouchaud}
\affiliation{Capital Fund Management \& Académie des Sciences}
\date{August 2023}
\begin{abstract}
This is the English version of my inaugural lecture at Collège de France in 2021, available \href{https://www.youtube.com/watch?v=bxktplKMhKU}{{\color{blue} here}}. I reflect on the difficulty of multi-disciplinary research, which often hinges of unexpected epistemological and methodological differences, for example about the scientific status of {\it models}. What is the purpose of a model? What are we ultimately trying to establish: rigorous theorems or {\it ad-hoc} calculation recipes; absolute truth, or heuristic representations of the world? I argue that the main contribution of  statistical physics to social and economic sciences is to make us realise that unexpected behaviour can emerge at the aggregate level, that isolated individuals would never experience. Crises, panics, opinion reversals, the spread of rumours or beliefs, fashion effects and the {\it zeitgeist}, but also the existence of money, lasting institutions, social norms and stable societies, must be understood in terms of collective belief and/or trust, self-sustained by interactions, or on the contrary, the rapid collapse of this belief or trust. The Appendix contains my opening remarks to the workshop ``More is Different'', as a tribute to Phil Anderson.  
\end{abstract}

\maketitle
{\it Volant super omnes}\footnote{The Collège de France has always represented for me a compendium of excellence and high standards, a unique link in the transmission of knowledge that is both cutting-edge and universal, instantaneous and permanent.  The courses given by Pierre-Gilles de Gennes, Claude Cohen-Tannoudji and Philippe Nozières, each in their own style, inspired and profoundly influenced me.  I don't think I would have become the researcher I am today without their teachings. That's why I'm so excited at the prospect of teaching here myself, if only for a year, and I'd like to thank the Bettencourt-Schuller Foundation warmly for giving me this opportunity, as well as my colleagues who proposed and defended my application, especially Antoine Georges and Alain Prochiantz. I'd also like to send many thanks: to Elisabeth, my lifelong muse, to my children and their stimulating ideas, to my colleagues, collaborators and friends, but also to my students and post-docs. After all, research is first and foremost a collective phenomenon.}

\section{Introduction}

I have chosen to devote my inaugural lecture to the challenges posed by multi-disciplinarity, in the hope that my experience as a theoretical physicist exploring economics and finance will illustrate the issues and pitfalls associated with an approach that is supposedly fruitful and encouraged by the supervisory bodies, but which comes up against numerous obstacles when we try to put it into practice. 

In many ways, venturing into a discipline that is not your own is like emigrating to a country where you know neither the language nor the culture, and which does not expect much from newcomers, especially when they have the arrogance and naivety to believe that they will bring a new perspective, different tools, a complementary representation of the world. And yet, as René Char said, {\it what comes into the world to disturb nothing deserves neither consideration nor patience.} 

Such confrontation of cultures, often difficult and sometimes brutal, is nevertheless fertile if it is allowed to continue over long periods. It also raises questions about science in general, and science in particular - what are its aims? What are its methodological presuppositions, its social conventions, its criteria of excellence and scientific truth, its editorial practices? Should theory precede confrontation with data, as some economists believe, or should observation inspire theory, as is often the case in physics? And what is a theory, a model, a law - terms which, curiously enough, do not have the same meaning or the same status in different disciplines, as I will come back to later. 

This discussion is particularly important because models often serve as paradigms, as conceptual pillars on which disciplines develop. Make no mistake: certain models and theories, when we are exposed to them, carve such a deep groove in our minds that they determine, in the long term, our conception of the world and our professional practices. 

However, an interesting question is raised from the outset: what legitimacy does one have to criticise a discipline to which one does not belong? Are lay critics credible, or even audible? Can one cast doubt on the knowledge of experts in their field by relying on expertise from outside that field?\footnote{This paradox echoes the current debate about the value of expert opinion in a democracy.}  And yet, a fresh perspective is sometimes needed to cast doubt on axioms that are too familiar or to question the obvious that is too consensual. 

As far as methodology is concerned, such interference seems to me to be perfectly legitimate, because the practice of scientific research and the production of knowledge can undoubtedly be transposed to other fields, beyond the purely technical aspects. It's not uncommon for the formal beauty of a theory, its mathematical aesthetics, to take precedence over its relevance to describing the world; in such cases, research develops outside the ground, in a vacuum, where the justification becomes the sheer intellectual investment already made. In this case, it is very difficult to break the spell; on this point, for example, I would like to quote the economist Willem Buiter, who wrote in 2008, shortly after the onset of the great recession:\footnote{Willem Buiter, The unfortunate uselessness of most 'state of the art' academic monetary economics. VoxEU (2009). https://voxeu.org/article/macroeconomics-crisisirrelevance.} {\it Most mainstream macroeconomic theoretical innovations since the 1970s have turned out to be self-referential, inward-looking distractions at best. Research tended to be motivated by the internal logic, intellectual sunk capital and aesthetic puzzles of established research programmes rather than by a powerful desire to understand how the economy works.}
   
\section{From Statistical Physics to Social Sciences}

I discovered statistical physics in the early 80s. One of the major research themes at that time was phase transitions (after its heyday in the 1970s), which brought to light the concepts of collective effects and universality, which had a scope far beyond physics alone, as I will discuss later.

A new body of theory was then being built around the somewhat vague notion of ``complex systems". It was discovered that some physical systems never reach thermodynamic equilibrium, because reaching that equilibrium would be like solving an optimisation problem so complex that no algorithm -- not even the dynamics of the system itself -- could find the solution in a reasonable amount of time.  We also understood that these systems are fragile, i.e. hyper-sensitive to disturbances, and their dynamics intermittent. Small causes can give rise to major catastrophes (avalanches, earthquakes), leading to abrupt and discontinuous evolution. The statistics of events, far from being Gaussian, are characterised by thick distribution tails, which allow extreme events to occur with an appreciable probability.  

A telling example of such situations is the anomalous diffusion of the Lévy type. We are all familiar with Brownian motion, in which diffusion results from a very large number of very small steps. Brownian motion is continuous, and the displacement statistics are Gaussian. In contrast, the ``Lévy flight" is made up of steps of different sizes, some very small and others very large. Whatever the scale at which we observe this movement, it is always the few largest jumps that dominate the total displacement. A Lévy flight is perfectly self-similar, although it is dominated by extreme events. 

Such anomalous diffusion was first observed experimentally in 1990 at the Laboratoire de Physique Statistique de l'ENS, in a solution of so-called 'giant' micelles \cite{ott1990anomalous}.  I was lucky enough to be able to interpret these experiments theoretically, and this work was one of the triggers for my gradual migration towards what is now called 'econophysics', a rather unfortunate neologism but constructed in the same way as 'biophysics' or 'geophysics'.

Because a few years earlier, in September 1987, the first conference bringing together economists and physicists was held in Santa Fe (New Mexico). Co-organised by Phil Anderson - in my view one of the most extraordinary theoretical physicists of the twentieth century (Nobel Prize in Physics in 1977) - Ken Arrow (Nobel Prize in Economics in 1972) and David Pines (Quantum liquid physicist), the conference was entitled ``The Economy as an Evolving Complex System" \cite{anderson2018economy} sought to imbue theoretical economics with recent ideas -- perhaps too recent -- from statistical physics.  

Unfortunately, despite such prestigious godfathers, the graft did not take. For reasons that I will try to explain later, econophysics developed for 30 years in a relatively autarkic way, creating very few bridges with the world of economists and a little more with the world of theoretical finance. 
Neo-classical economics seemed to have won the day; in 2003 -- 5 years before the great recession of 2008 -- Robert Lucas, winner of the 1995 Nobel Prize in Economics, declared: {\it macroeconomics [...] has succeeded: its central problem of depression prevention has been solved, for all practical purposes}.

And yet, ironically, on 19 October 1987, just a few weeks after the Santa Fe conference, the Dow-Jones index experienced its worst day ever, falling 22.6\% in a single day. A fine example of a rare event, totally absent from the classical theory of supposedly Gaussian markets. What's more, this crash does not seem to be linked to any economic news that might explain its scale.

Faced with such an event with no apparent cause, the physicist immediately thinks of an endogenous discontinuity, generated by feedback loops internal to the complex system that is a financial market, with its thousands of players who influence each other... but this interpretation was a long way from the dominant dogma at the time (or even now), that of rational agents, whose infinite wisdom would allow the markets to always display the 'true' price, the one that reflects the so-called 'fundamental' value, and that only varies if new information justifies it. This is the theory of efficient, stable markets, which do not get carried away spontaneously, do not give in to unjustified panic movements, but stubbornly provide information about the value of the world. 

{\it Do you guys really believe this?} was Phil Anderson's blunt and undiplomatic reaction when this theory was presented to him in Santa Fe in 1987. I admit I had the same reaction when, a few years later, I read my first theoretical finance articles, particularly those on the famous Black, Scholes \& Merton model. This seems to me to be a perfect example of the gulf that can exist between scientific cultures. It allows me to discuss in detail to the essential differences between possible interpretations of the very concept of 'model', and to discuss a crucial question: {\it what is a good model?}  

\section{A Paradigmatic Model: Black, Scholes \& Merton}

After Fisher Black's untimely death in 1995, Myron Scholes and Robert Merton were awarded the Nobel Prize in Economics in 1997 for their theory of option pricing \cite{black1973pricing}, which I have to present here very briefly so that its paradigmatic nature can be understood.

An option is an insurance contract on the value of a financial asset. Suppose you want to insure your share portfolio so that even if prices fall significantly, the insurer undertakes to buy it back at a minimum price, say today's price, for a period of ten years. What premium is the insurer entitled to charge? Between now and -- say -- one year from now, what investment strategy should the insurer follow to minimise the risk it is covering? 

The Black-Scholes-Merton model makes it possible to answer these two questions simultaneously and unequivocally. There are two surprises, which are particularly relevant to my point, because they offend the intuition of the physicist, but seem natural to the economist and to the mathematician. What is -- or has become -- obvious to some may remain obscure or even questionable to others. These discrepancies can either give rise to new ideas or, on the contrary, breed mutual distrust.  

The first surprise is that, in the world of Black-Scholes, the insurance premium does not depend on the value of the expected average return of the stock market over the next year; whether the stock market rises or falls, the price of the options is unchanged; no assumption about this average return is therefore necessary to set the price of insurance against falling stock prices! 

The second, perhaps even more disturbing surprise is that the insurer can follow a strategy known as 'hedging', which allows it to take {\it no risk at all}.  However, the model explicitly assumes that market prices are unpredictable.  By what sleight of hand, then, can risk disappear completely?

The answer lies in the very special nature of the Black-Scholes-Merton hypotheses, which assume that price movements are described by a continuous-time Brownian motion. As I mentioned earlier, Brownian motion can be constructed as the superposition of an infinite number of infinitesimally small steps. In such a model, the price evolves continuously, without jumps; chance is ``benign", uncertainty can be tamed and, in a sense, disappears. 

But it is a singular, fragile model: as soon as one takes into account the fact that the statistics of price variations are heavy-tailed, that price jumps of all sizes occur in financial markets\footnote{Earthquakes of very different magnitudes coexist, some imperceptible and others destructive. Remember that the Richter scale is logarithmic: +1 on this scale corresponds to an earthquake 30 times more powerful.}, from the insignificant to century-long crashes like the one in 1987, then the risk that Black and Scholes had swept under the carpet reappears, and it is in fact far from negligible. Moreover, this risk is asymmetrical: the insurer's potential losses are much greater than those of the insured. If this were not the case, the very existence of insurance contracts of this type would be pointless, since insurance, by definition, is a transfer of risk. In a way, the model contradicts the very existence of the contracts it seeks to theorise. 

But it gets even worse. The indiscriminate use of the Black-Scholes-Merton model creates a destabilising feedback loop, which considerably amplified the 1987 crash I mentioned earlier. So the paradox is complete: using a model in which crashes do not exist can actually trigger their existence!  The model itself becomes a systemic risk factor.

My first foray into theoretical finance owes a great deal to Christian Walter, who had read my article on Lévy flights in giant micelles \cite{ott1990anomalous}. He suggested that I should generalise the Black-Scholes-Merton model to situations of 'wild randomness', as Benoît Mandelbrot called it, i.e. where the randomness is heavy-tailed. The result of this research lead to an article published in Le Monde, March 14$^{th}$, 1995, entitled {\it Les marchés dérivés, pour une pédagogie du risque}. 

The least that can be said is that my piece did not help to establish a calm dialogue with financial mathematics. But it did lead to the creation of Science \& Finance, which later merged with Capital Fund Management. This enabled us to develop an alternative, 'physicist's view' of financial markets and, more generally, of theoretical economics. Looking back, I still do not know whether a less confrontational approach would have accelerated the innovation process, or whether it would have diluted or even stifled original ideas. That's precisely the conundrum of multi-disciplinarity.   
  
\section{Phenomenological Models}

The Black-Scholes-Merton model belongs to a category that I would call 'phenomenological' models, to be distinguished from 'foundational' and 'metaphorical' models, which I will discuss later. The polysemy of the word 'model' is in fact a first obstacle to multi-disciplinarity: what is the purpose of a model? What makes it scientific? What are we ultimately trying to establish: rigorous theorems or ad-hoc calculation recipes; absolute truth, or heuristic representations of the world?

``Phenomenological" models attempt to represent reality in mathematical terms. Using a few formulas or 'laws', they summarise the behaviour they are trying to describe, for example a simple linear regression. The mathematical apparatus then takes these laws and draws all sorts of logical consequences from them. From being descriptive, the model then becomes predictive. 

An example of this type of model is given by the law of behaviour of so-called non-Newtonian fluids, such as mayonnaise, maizena or toothpaste, which stipulates that the stress $\sigma$ applied is equal to a threshold stress $\sigma_c$, increased by a power law contribution from the shear rate $\dot \gamma$: 
\begin{equation}
   \sigma = \sigma_c + \dot \gamma^x.
   \label{eq:1}
\end{equation} By injecting this behavioural law into the general equations describing the flow of liquids, we can predict, for example, the head losses in a pipe, i.e. the cost of pumping such a fluid. We therefore obtain a prediction of the model, but no explanation of the origin of the postulated law of behaviour. 

As René Thom once noted, to {\it predict} does not necessarily mean to {\it understand}. The law used is at best justified by empirical observations that are adjusted, within a certain range of variability, by the law in question. 

But these laws are often motivated only by considerations of mathematical convenience. The phenomenological model always contains a certain number of adjustable parameters, which are chosen to represent the phenomenon ``as well as possible". It is thus {\it always} possible to find the optimum value for these parameters, even if the model is totally unrealistic. But if the model is wrong, such a procedure can be disastrous, because the consequences deduced from the model may be completely at odds with reality. 

There are several models widely used in finance which are precisely in this situation: one can adjust the observations {\it perfectly} on a given day, calculate the consequences {\it exactly}, but end up being completely wrong about the sign of certain derivatives, i.e. predict that certain quantities should increase when in reality they are decreasing!\footnote{The so-called ``local volatility'' model is one example.}

The problem is that there is often no justification for these models based on {\it fundamental principles}. These principles are the subject of a consensus because all the consequences drawn from them are consistent with all known observations -- as is the case, for example, of thermodynamics or quantum mechanics. To return to non-Newtonian liquids, we would like to justify the rheological law, Eq. \eqref{eq:1}, based entirely on the movement of the molecules that make them up, which is governed by quantum mechanics. What is the collective mechanism by which a threshold constraint $\sigma_c$ arises, preventing the liquid from flowing if the constraint is not strong enough? This is the heart of statistical physics, and one of the themes of my forthcoming lectures: how can we move from the microscopic to the macroscopic? 

The reason why such a ``microscopic" justification is essential is that it allows us to understand the limits of a model, and when it is pushed outside the normal operating regime. What happens in extreme situations? Is the model still valid? This is obviously an essential concern, both in conventional engineering and in financial engineering; and also, for example, for the monetary policy of central banks in times of crisis.\footnote{During the workshop ``More is Different'' at the end of my lectures, Jean-Claude Trichet will talk about the difficulties he encountered when he was President of the European Central Bank during the 2008 crisis, which led him to say: {\it Models failed to predict the crisis and seemed incapable of explaining what was happening [...] in the face of the crisis, we felt abandoned by conventional tools.}}  

The Black-Scholes-Merton model is a perfect case in point. We postulate an intellectually reasonable model (that of the Brownian motion) for which the mathematical tools are elegant, efficient and plentiful, and we turn the crank that allows us to calculate all sorts of things, while turning a blind eye to the manifestly non-Gaussian events (at the beginning of the 90s, it was even accepted that one should remove the so-called aberrant events from the calibration!) This established a paradigm of zero risk, which led to the crash of 1987 that contradicted its very premises. 

As I said earlier, the importance of paradigms should not be underestimated: they inform our intuition and shape our vision of the world. The dangers of a model without 'microscopic' foundations are magnified when it is performative, i.e. when it establishes a certain number of professional or institutional practices based on the model itself.  

In the case of the Black-Scholes-Merton model, a missing but essential ingredient is the feedback of the hedging strategy on prices. The price of a financial asset does not exist in itself, as the Platonic vision of efficient markets would have us believe (and as financial engineers have long believed, and sometimes still believe). The price is but the result of the very actions of buyers and sellers. Transactions have an impact on prices, and this impact can initiate a feedback loop and lead to instability (see for example \cite{bouchaud2011endogenous}).  It is by trying to formulate a more 'microscopic' theory of price formation, instead of postulating an ex-nihilo model, that we might have guessed the intrinsic limitations of the Black-Scholes-Merton model. 

In any case, it seems to me crucial that the qualitative conclusions of a model should be robust to small changes in assumptions, because reality rarely obeys perfectly the equations we postulate. A model is a simplified representation of the world that allows us to orientate ourselves and guide our intuition. The map is not the territory, and we need to make sure that the simplification is not a betrayal. A good model should be a non-paradoxical representation of the world: there should be as few anomalies, paradoxes, exceptions and 'black swans' as possible.

\section{Fundamental Theories and 'Metaphorical' Models}

Phenomenological models coexist with, and often feed into, two other types of model whose aims are different.\footnote{The classification proposed here is neither exhaustive nor exclusive. There are other types of model, such as those based exclusively on machine learning, which bypasses the classic stages of theoretical modelling. Nor, in my mind, is there any Comte-like hierarchy between these different types of model: their objectives are simply different and complementary, and their relevance depends on the context in which they are used.} ``Foundational" models are more ambitious: they do not propose {\it ad hoc} equations to describe a phenomenon, but rather an axiomatic, explanatory and coherent scheme. If it stands up to various logical and empirical tests, such a model becomes a fundamental theory. A good example is the Bohr-Rutherford model of the hydrogen atom: it postulates that the atom is made up of a positively charged nucleus and an electron constrained to move in stable orbits, determined on the basis of a condition for quantifying angular momentum. This model contains the premises of quantum mechanics. 

Another example is the rational agent model in economics, which is the starting point for an ambitious theory of human decisions in the presence of uncertainty, from which all sorts of economic conclusions can be drawn. Unfortunately, in this case, the experimental evidence from numerous cognitive and behavioural biases on the part of economic agents casts strong doubts on the validity of the initial model -- unless we consider that at the aggregate scale of an economy (or a financial market), irrational individual behaviour is averaged out and disappears. Everything would then happen ``as if" agents were rational. But is it really reasonable to make this additional assumption? 

In any case, this once again concerns the question of the path from micro to macro: what remains of individual behaviour when we observe large ensembles? We shall see that in the presence of interactions these biases may, on the contrary, be reinforced rather than average out to zero. 

Finally, I would like to talk about some models that are particularly close to my heart, which I propose to call 'metaphorical'; they are also often called 'toy-models'. These models do not seek to explain an observation in detail, but rather to highlight {\it mechanisms} that would be difficult to identify without a precise mathematical framework. The hypotheses of these models are crudely stylised, sometimes even seemingly absurd, but their consequences are non-trivial, difficult to guess without the model, and seem to account for a real phenomenon. 

The metaphorical model is a catalyst for the imagination, a crutch for thought. It helps us think about the phenomenon by providing an explanatory {\it scenario}. It suggests other observations, other experiments, which, depending on the case, will either reject or validate the proposed mechanism. The metaphorical model can then gradually be enriched and eventually become usable in practice.  

But in its primitive version, the metaphorical model is not intended to be 'calibrated' on experience; it is a conceptual proposition which must first be confirmed qualitatively, before being confirmed quantitatively, on the basis of the details of reality. 

Such gradual approach is common in some disciplines, but strongly rejected in others. In particular, the absence of confrontation with reality is sometimes perceived as an admission of weakness, whereas in reality it is a desire not to rush through the stages: we want first to be sure that we have identified the relevant mechanisms before trying to adjust the data at all costs, even if it may mean abandoning the idea halfway through. Without the freedom to make mistakes, there can be no fruitful research.

There are many examples of metaphorical models that have become paradigmatic. One example is the Hopfield model \cite{hopfield1982neural}, which provides an extremely stylised way of understanding how an assembly of neurons can learn to memorise information, whereas each individual neuron is, of course, incapable of doing so. But the 'neurons' in Hopfield's model are such caricatures that it is difficult for a specialist in nerve cells to accept that such a representation could even be relevant. 
Hopfield's model is therefore not intended to be 'calibrated' on neurophysiological data. However, Hopfield's initial article has over 26,000 citations (as of 2023), and it  founded a new discipline, computational neuroscience, which is producing increasingly realistic models of how the brain works. 

Another famous example is the Wigner model of atomic nuclei with a large number of nucleons \cite{wigner1967random}. The Hamiltonian of these systems is too complex for an analytical calculation of the energy levels to be possible. Eugene Wigner therefore proposed replacing the kinetic energy and interaction energy of the nucleons with a totally random matrix, in which each element is chosen independently, while respecting the symmetry of the initial problem. Unexpectedly and quite remarkably, the statistics of the eigenvalues of these random matrices reproduce very precisely the statistics of the energy spectrum of heavy nuclei and, more generally, of chaotic quantum systems.

Wigner's radical idea, based on intuition rather than deductive reasoning, led to the development of a particularly fascinating branch of probability and theoretical physics: Random Matrix Theory, which today has a great many practical applications \cite{potters2020first}.   

Wigner's work has also been emulated in other fields, such as May's model of ecological systems \cite{may1972will}. Robert May assumes that species interact randomly with each other, either in competition or collaboration. May shows that the system becomes generically unstable when the number of species increases, whereas it was thought that high ecological diversity favoured stability. (Similarly, it was thought that a greater dispersion of banking risks favoured the stability of the financial system, whereas this dispersion also favoured the contagion of a loss of confidence, as happened in 2008). May's model was published in 1972 in a one-and-a-half page article entitled ``Will a complex system be stable" \cite{may1972will}; the assumptions are caricatured, but he too launched a whole discipline, that of theoretical ecology.

Bernard Derrida's Random Energy Model \cite{derrida1981random} also assumes that the energy of a configuration of molecules in a liquid is a random variable, chosen independently for each configuration. This is an almost absurd assumption, since if the position of just one of these molecules is changed, the energy of the resulting configuration will be very close to that of the initial configuration. However, Derrida's model predicts a glass transition whose characteristics closely resemble those observed experimentally. Derrida's 1980 paper is considered to be the precursor of the modern theory of structural glasses, known as the ``RFOT" theory (Random First Order Transition), and has also found applications in a wide range of fields (see e.g. \cite{biroli2012random}).

\section{From the Individual to the Collective}

There are many more examples of this type, but I would like to mention in particular a metaphorical model invented by Thomas Schelling, winner of the Nobel Prize in Economics in 2005, in an attempt to explain the phenomena of urban segregation.

Schelling's work is particularly important for the subject I have chosen to deal with today and in the weeks to come. His book, 'Micromotives and Macrobehaviour' \cite{schelling2006micromotives}, whose title could have been chosen by a physicist, deals precisely with this path from micro to macro  that I have already mentioned several times, and which seems to me to be the major contribution of statistical physics to the social sciences. 

Schelling is regarded as the precursor of agent based models in economics.\footnote{Agent based models are intended to be foundational, but their complexity is such that they need to be simplified in order to dissect and understand the mechanisms at work. In this sense, they become "metaphorical".} These models attempt to reconstruct the macroeconomy, the economy of aggregates, on the basis of a very large number of imperfect, heterogeneous agents, all interacting with one another. Interaction here means that the decisions of each agent are directly influenced by what the others say and do, whether their analysis is rational or not. These interactions create instabilities, phase transitions and discontinuities, which I believe are highly relevant to understanding economic systems and financial markets, but which are still not properly taken into account in current macroeconomic models. 

Before returning to this essential theme, what is Schelling trying to describe with his caricatured model? He assumes that every individual prefers to live in a moderately dense neighbourhood, because the quality of life is better there than in an overcrowded or deserted neighbourhood. The total population is such that moderate density is theoretically achievable in all neighbourhoods, in other words, all individuals can in principle be satisfied at the same time (which is of course not always the case in real-life situations).

Each resident can choose to move if they visit a neighbourhood that seems closer to the optimum density than the one they are presently living in. The result of these individual decisions is unexpected: after a while, an equilibrium is created in which some neighbourhoods are overpopulated and others completely deserted, instead of all being optimally populated. 

As we can see, this result is totally counter-intuitive. This ``unwitting" segregation is a consequence of the agents' egocentric behaviour, and can be demonstrated rigorously using an analogy with the liquid-gas transition, much studied in statistical physics \cite{grauwin2009competition}. Schelling's segregation phenomenon is similar to a classical liquid-gas phase separation.

Schelling therefore constructed a counter-example to Adam Smith's invisible hand: it is not because individuals follow their own interests that a state of optimal collective well-being is achieved. In some cases, the opposite is true!  Contrary to the intuition built into classical economic (or political economy) models, an assembly of rational individuals can behave irrationally.    

Schelling's model, like Hopfield's, illustrates a very important phenomenon, which has in a way become the major trophy of statistical physics: totally unexpected behaviour can appear when we observe groups of interacting individuals. This is what Phil Anderson emphasised in his famous 1972 article entitled ``More is Different" \cite{anderson1972more}: collective behaviour cannot be understood as a simple superimposition of individual behaviour.\footnote{{\it The behavior of large and complex aggregates of individual elements is not a simple extrapolation of the properties of a few elements. Instead, {\bf entirely new properties appear}, which require specific models. See also Appendix. }} In particular, even when individual behaviour evolves continuously, collective behaviour can be discontinuous. 

A classic example is that of phase transitions: although the same $H_2 O$ molecules exist on a microscopic scale, ice at a temperature of $0^{\circ -}$ and liquid water at $0^{\circ +}$ behave completely differently on a macroscopic scale. Liquid water is blue and flows, ice water is white and rigid. It is a phenomenon so familiar that we forget its counter-intuitive, almost miraculous nature: interactions between molecules can lead to truly unexpected emergent behaviours, such as superfluidity or superconductivity, which are truly spectacular collective effects. 

This is where metaphorical models are particularly illuminating: they allow us to grasp phenomena that our imagination has great difficulty in conceiving. For example, memory (like consciousness) is a collective property, which does not exist at the level of individual neurons, and which Hopfield's model helps us to understand. 

Collective phenomena of this kind also occur spontaneously in animal and human communities: the extraordinary figures of swarms of starlings \cite{ballerini2008interaction} or schools of fish, the spectacular synchronisation of the flickering of thousands of fireflies,\footnote{see, e.g. Strogatz, S. (2004). Sync: The emerging science of spontaneous order. Penguin UK.: {\it How can thousands of fireflies orchestrate their flashing so precisely and on such vast scales? For decades, no one could come up with a plausible theory. A few believed there must be a maestro, a firefly that cues all the rest.}} the ``Ola" in stadiums, collective psychogenic syndromes, but also crowd movements, such as the murderous ones in Mecca.

In the latter case, it can be shown that the crowd is so dense that individuals disappear and merge into a continuous human matter, a genuine elastic medium in which pressure waves can propagate \cite{johansson2008crowd}. This emergent phenomenon does not exist if the density is too low: in this case, the movement of an individual affects only a few neighbours. Waves only propagate beyond a threshold density, known as the {\it jamming transition}, which has been the subject of a great deal of study by physicists over the last twenty years \cite{charbonneau2017glass}, and which is closely linked to the appearance of a threshold stress in non-Newtonian fluids, which I mentioned earlier (see Eq. \eqref{eq:1}). 

In fact, it is these pressure waves that are lethal, because they can focus and compress people in such an extreme way that they cannot escape. Individuals no longer exist as such: they propagate these waves in spite of themselves, but it is their individual movements that generate the waves. These waves cannot be reduced to the simple sum of elementary effects; they are fundamentally {\it something else}. The crowd is not an extrapolation of isolated individuals.

This transubstantiation of the individual into the collective is, in my view, a fundamental ingredient in understanding a range of socio-economic phenomena, which make no sense when we try to explain them on the basis of the behaviour of isolated individuals. On the contrary, the effects of {\it imitation} and {\it contagion} seem essential, for better or for worse.\footnote{The role of narrative narratives in propagating opinions seems essential, as Robert Shiller illustrates in his latest book: R. Shiller, "Narrative Economics", Princeton University Press, 2019.} Economic, financial or social crises, banking panics, opinion reversals, the spread of rumours or beliefs, fashion effects and the {\it zeitgeist}, but also the existence of money, lasting institutions, social norms and stable societies, must in my view be understood in terms of collective belief and/or trust, self-sustained by interactions, or on the contrary, the rapid collapse of this belief or trust. How else, for example, can we understand the sudden disappearance of certain political parties after decades of hegemony, or the radical changes in cultural representations, as we have seen again recently? 

Like Theseus' ship, which retains its identity even as all its parts are gradually replaced, it is because culture, customs and conventions belong to the collective and not to individuals that they can be passed on and survive when the individuals themselves have disappeared. But precisely because of these collective effects, yesterday's lore and consensus is sometimes brutally challenged and become today's heresy.

In recent metaphorical models, collective belief, trust or mistrust are emergent phenomena that occur within a limited range of parameters, just as the range in which water is ice is limited \cite{bouchaud2013crises}. Small variations in socio-economic conditions (or in the perception of them) are enough to tip the system into a crisis, which translates into the destruction of collective structures. These metaphorical models have the virtue of making us aware of the fragility of social constructs, which are essential to our prosperity and which we tend to take for granted. But our economic and social environment is not determined exogenously; it is shaped by consensus, whether spontaneous or imposed by laws and institutions.  Just as ice melts when heated above $0^\circ$, these collective syntaxes can implode suddenly, without warning, when certain conditions that we still know little about are met.

The 2008 crisis, for example, was interpreted by many observers -- including Ben Bernanke, then Chairman of the Fed \cite{bernanke2019firefighting} -- as a collapse in collective confidence in the financial system, with no 'objective' material cause to justify its scale, just like the 1987 crash I mentioned earlier.\footnote{As early as the 19th century, Walter Bagehot observed: {\it The peculiar essence of our financial system is an unprecedented trust between man and man; when that trust is weakened by hidden causes, a small accident may greatly hurt it}. W. Bagehot, Lombard Street (1873)} In his inaugural speech in January 2009, Barack Obama admitted his perplexity and dismay at the brutality of the upheaval: {\it Our workers are no less productive than when this crisis began. Our minds are no less inventive, our goods and services no less needed than they were last week...}

But the success of public policies is often a matter of social engineering -- how do you get people on board, re-establish trust, avoid panic? I believe it is important to understand the mechanisms underlying these collective phenomena, particularly in these times of economic, financial and, more recently, health crises, with social networks that are even more conducive to self-amplifying effects and the irrationality of crowds.

\section{Conclusion}

The phase transition paradigm therefore makes it possible to account for such discontinuities, for catastrophic effects without commensurable causes, and for endogenous instabilities rather than those due to exogenous events.\footnote{{\it What shocks are responsible for economic fluctuations? Despite at least two hundred years in which economists have observed fluctuations in economic activity, we still are not sure.} Cochrane, J. H. (1994, December). Shocks. In Carnegie-Rochester Conference series on public policy (Vol. 41, pp. 295-364). North-Holland.} These instabilities are often the direct consequence of a certain form of optimality, and this seems to me to be an important lesson: the search for an optimum often leads to a complex system becoming fragile.\footnote{On this point, see the very recent preprint: Moran, J., Pijpers, F. P., Weitzel, U., Bouchaud, J. P., \& Panja, D. (2023). Temporal criticality in socio-technical systems. arXiv preprint arXiv:2307.03546.} These are the avenues I intend to follow in the sessions to come and also during the lecture that will close my lectures, which is entitled 'More is Different', in homage to Anderson's article (see Appendix). 

In fact, we can find precursors to Anderson's thesis, particularly in the writings of John Maynard Keynes, whose insights and depth of thought never cease to impress me. Keynes had already clearly anticipated the importance of collective effects, fragility and discontinuity when he wrote: {\it we are faced [...] with the problems of Organic Unity, of Discreteness, of Discontinuity, the whole is not equal to the sum of the parts, [...], small changes produce large effects, the assumptions of a uniform and homogeneous continuum are not satisfied}. For Keynes, ``organic" means collective, as opposed to ``atomic" which refers to a simple superposition of individual processes, without the emergence of new effects at the aggregate level.  

So it is on the shoulders of giants that I make my case for a fertile synthesis between statistical physics and the social sciences. But to return to my introductory remarks, multi-disciplinarity is not necessarily self-evident. I would like to quote a recent article by Andy Haldane, at the time Chief Economist at the Bank of England \cite{haldane2019drawing}. In a paragraph entitled ``How did macroeconomics become insular?", he writes: {\it When asked in 2006 to agree or disagree with the statement 'In general, interdisciplinary knowledge is better than knowledge obtained by a single discipline', close to 60 per cent of academic economists said that they strongly disagreed.}

Dialogue between physicists and economists therefore remains difficult, even if it has improved considerably over the last twenty-five years, and particularly since the 2008 crisis, which brought systemic instabilities back to the fore. Many links, both formal and informal, have been established, involving prestigious institutions such as the OECD and the Bank of England. 

I hope that these lectures will provide an opportunity to strengthen these links, to give rise to innovative research projects -- since this is the title of the chair I have been given this year -- and to invent new models, whatever their nature -- phenomenological, metaphorical, foundational -- as long as they enable us to better understand emerging social phenomena, if only to try to prevent certain disasters of the past from happening again, or to better manage those that will undoubtedly 
happen again.

\section*{Appendix. ``More is Different'' : Opening Remarks}

My lectures at Collège de France were rounded by a two-day workshop entitled ``More is Different'', which gathered physicists, economists and mathematicians around the theme of crises and collective phenomena.\footnote{For the full program, see \href{https://www.college-de-france.fr/fr/agenda/colloque/more-is-different}{{\color{blue} here}}} Here are my brief introductory remarks to what was (I believe) a memorable event. 

\begin{center}
    $\bullet \bullet \bullet$
\end{center} 

Welcome to “More Is Different”, which I wanted to be both a tribute to Phil Anderson and his eponym 1972 paper \cite{anderson1972more} which impressed so many people, including myself, and a testimony of what I tried to convey in my lectures: the importance of emergent, collective effects, and the importance of multi-disciplinarity, hence the motley crowd of speakers invited to speak today and tomorrow: physicists, mathematicians, biologists, economists, computer scientists. 

A few words about Phil Anderson, who died in March 2020, just after this workshop was decided. Although less well known by the general public than scientific geniuses of the XXth century like Einstein, Feynman, Turing and many others, Phil Anderson is a true intellectual hero, and has influenced my way of doing and thinking about science ever since I started my PhD and discovered one of his main claim to fame: Anderson’s localisation, which led (at least in part) to his 1977 Nobel Prize. I don’t think it is an exaggeration to say that he could have won at least two other Nobel prizes, one for the Josephson effect, and one for the Higgs boson, also called the Anderson-Higgs mechanism. For those who want to know more about the scientific contributions and personal life of Phil Anderson, I strongly recommend the very good recent book of Andrew Zangwill called “A Mind Over Matter” (Oxford University Press), from which I borrowed some of the material below.

The breadth of his scientific interests was really amazing, as he described himself in a debate with particle physicist Steven Weinberg in 1991: “How did life begin? How does the brain work? What is the theory of the immune system?  Is there a science of economics?”, all these themes, you will have noticed, being clearly related to the present workshop. 

And he went on to say, along the lines he was already arguing in “More Is Different”: {\it All these things have in common that they are manifestations not of the elementary constituents of matter, but of the complex organisation of that matter.} One of his constant fight was to show that understanding complex organisations was as fundamental as understanding the elementary laws and constituents. In fact, he argued against the idea that a deep understanding of the building bricks could help understanding questions at the aggregate level. He wrote that {\it the ability to reduce everything to simple fundamental laws does not imply the ability to start with these laws and reconstruct the Universe}. 

In More Is Different, he introduces an even deeper idea, which strongly resonates with my own enthusiasm for complex systems – that of emerging surprises: {\it The behavior of large assemblies of interacting individuals (particles) cannot be understood as a simple extrapolation of the properties of isolated individuals (particles). Instead, entirely new, unanticipated behaviors may appear and their understanding requires new ideas and methods.} Although Anderson was not aware of it, this idea of emergence was stated in almost the same terms 50 years before by English philosopher C.D. Broad: {\it Emergence is the fact that the whole cannot be deduced from the most complete knowledge of its components}. 

There is perhaps an even more interesting twist to this idea, in particular in the context of social sciences and economics. Anderson offered a specific mechanism capable of producing emergence, i.e. unforeseen, and sometimes unimaginable aggregate properties: “symmetry breaking”, in a broad sense, leads to distinct macroscopic phases, i.e. regions of the parameter space where aggregate behaviour is qualitatively the same, but markedly different from the behaviour in other regions of the parameter space. Think of liquid water and ice for example. 

But here is the twist: not only emergent properties cannot be predicted from elementary building blocks: the same aggregate, macro properties appear for a very wide range of elementary, micro constituents. Only a handful crucial micro-properties seem to matter. The rest disappears upon aggregation, which can only produce a few types of macro-properties. This is actually good news for modelling: one does not need to nail down the exact behaviour of individual molecules, of individual neurons, or of individual economic agents to reproduce the qualititative behaviour of the very many. Only a rough sketch of these elementary ingredients may suffice, while still generating surprising features at the macro level. 

Phil Anderson was of course confident that these ideas might be relevant for economic sciences when he accepted to co-organise the first “econophysics” meeting at the Santa Fe Institute in 1987 (although the name econophysics was not yet minted).  The title of the meeting was telling: “The Economy as a Complex Evolving System” \cite{anderson2018economy}. Although this was the first of a series of three, with lots of great ideas put forward, I think it is fair to say it had up to now only limited influence on mainstream economics. Although things are slowly changing, I have heard economists say: been there, done that, all this complexity thing was just a hype, we moved on. 

I, together with many others like Alan Kirman, disagree, I think the best contributions of complex systems to economic and social sciences are yet to come. At least this was my hope when I gave my lectures here at College de France, and it is again my hope gathering here for only two days among the most brilliant minds working on these topics. 

Maybe what was missing in the nineties was a deeper involvement of physicists in this endeavour. It won’t do to just say – look at these ideas, great stuff, they should help you solving your problems. True, long term multidisciplinary projects are needed. Perhaps the discovery of High Temperature Superconductivity in 1986 is, to some extent, responsible for the relative failure of such a collaboration. It so happened that Phil Anderson was so excited about them, and was so convinced that he had found the right explanation, that he just let go the Santa Fe project. If he had stayed involved longer, I am quite certain that he would have, yet again, invented a simple but highly relevant toy model, probably luring a much larger number of theoretical physicists to these topics. 

\twocolumngrid
\bibliographystyle{unsrt}
\bibliography{bibs}

\end{document}